\documentclass[twocolumn,showpacs,preprintnumbers,amsmath,amssymb]{revtex4}
\usepackage{amsmath}
\usepackage{graphicx}% Include figure files
\usepackage{dcolumn}% Align table columns on decimal point
\usepackage{bm}% bold math
\usepackage{mathrsfs}
\usepackage{graphicx}
\usepackage{float}
\begin{document}

%\preprint{APS/123-QED}
%\begin{CJK*}{GBK}{song}
\title{ Multiple teleportation  via the partially entangled states }

\author{Meiyu Wang, Fengli Yan}

\email{flyan@mail.hebtu.edu.cn}

\affiliation {~\\
    College of Physics Science and Information Engineering, Hebei Normal University, Shijiazhuang 050016, China\\
   ~
Hebei Advanced Thin Films Laboratory,  Shijiazhuang 050016, China}

\date{\today}

\begin{abstract}
{We investigate the multiple teleportation with some nonmaximally
entangled channels.  The efficiencies of two multiple teleportation
protocols, the separate multiple teleportation protocol (SMTP) and
the global multiple teleportation protocol (GMTP), are calculated.
We show that GMTP is more efficient than SMTP.
 }
\end{abstract}
%\end{CJK*}

\maketitle

%\section{Introduction}
Quantum teleportation \cite {s1} is one of the most significant
components in quantum information processing, which allows indirect
transmission of quantum information between distant parties by using
previously shared entanglement and classical communication between
them. Indeed, it is considered as a basic building block of quantum
communication nowadays. Not only is it one of the most intriguing
phenomena in the quantum world, but also a very useful tool to
perform various tasks in quantum information processing and quantum
computing \cite {s14,s15}. For example, controlled quantum gates are
implemented by means of quantum teleportation, which is very
important in linear optical quantum computation \cite {s2, s3}.
Recently, the original scheme for teleporting a qubit has been
widely generalized in many different ways  \cite {s5, s6, s7, s9,
s10, s11,s12,s13, s16}. In the previous teleportation protocols and
in many other applications of teleportation, we want to construct an
unknown input state with unity fidelity at another location while
destroying the original copy, which is always achieved if two
parties share a maximally entangled state. However, it might happen
that our parties do not share a maximally entangled state. This
limitation can be overcome by distilling out of an ensemble of
partially entangled states a maximally entangled one \cite {s4}. But
this approach requires a large amount of copies of partially
entangled states to succeed. Another way to achieve unity fidelity
teleportation with limited resources is based on the probabilistic
quantum teleportation protocols of Refs. \cite {s5, s6, s7}.

Recently, in an interesting work, Mod{\l}awska and Grudka \cite {s8}
showed that if the qubit is teleported several times via some
nonmaximally entangled states, then the \textquotedblleft errors"
introduced in the previous teleportations can be corrected by the
\textquotedblleft errors" introduced in the following
teleportations.
 Their strategy was developed in the
framework of the  scheme proposed in Ref.\cite {s3} for linear
optical teleportation. In this paper, we show that this feature of
the multiple teleportation of Ref.\cite {s8} is not restricted to
the teleportation scheme stated in Ref.\cite {s3}. Based on the
general teleportation language of the original  proposal  shown in
Ref.\cite {s1}, we compare the efficiencies of two multiple
teleportation protocols, the separate multiple teleportation and the
global multiple teleportation. In the former protocol, a complete
teleportation including error correction is strictly executed by
neighboring parties. On the other hand, in the latter protocol, all
errors introduced in the teleportation are corrected by the final
receiver. We find the global multiple teleportation is more
efficient than the separate multiple teleportation.

 To illustrate two
protocols clearly, let us first begin with the multiple
teleportation in the case of three parties.

Alice wants to teleport an unknown quantum state
\begin{equation}\label{2}
    |\psi\rangle=a|0\rangle+b|1\rangle
\end{equation}
to Bob, where $a, b \in C$ and $|a|^2+|b|^2=1$. There is no direct
entanglment resource between Alice and Bob, fortunately, Alice and
the third party Charlie have a partially entangled state
\begin{equation}\label{1}
    |\Psi\rangle=\alpha|00\rangle+\beta|11\rangle,
\end{equation}
while Charlie and Bob share the same entanglment resource, where
$\alpha$ and $\beta$ are real numbers and satisfy
$\alpha^2+\beta^2=1$. Without loss of generality, we suppose
$|\alpha|\leq |\beta|$.

The simplest and directest strategy is to perform two  separate
teleportations, i.e., Alice teleports the quantum state
$|\psi\rangle$ to Charlie via the first teleportation. Then Charlie
teleports it to Bob via the second teleportation. Because this
procotol consists of two separate teleportations, we call it the
separate multiple teleportation procotol (SMTP).

 According to the standard probabilistic
teleportation protocol, in the first separate teleportation, Alice
performs the Bell-basis measurement (BM) on the teleported qubit and
the entangled qubit in her side. Charlie can apply the corresponding
Pauli transformation conditioned on the result of BM, i.e., $I$ if
the BM yields $|\Phi^+\rangle$, $\sigma_z$ for $|\Phi^-\rangle$,
$\sigma_x$ for $|\Psi^+\rangle$, and ${\rm i}\sigma_y$ for
$|\Psi^-\rangle$ , where $I$ is the identity, $\sigma_x, \sigma_y,
\sigma_z$  are standard Pauli matrices and
\begin{equation}\label{BM}
|\Phi^\pm\rangle=\frac{1}{\sqrt{2}}\left(|00\rangle\pm|11\rangle\right),
\end{equation}
\begin{equation}
|\Psi^\pm\rangle=\frac{1}{\sqrt{2}}\left(|01\rangle\pm|10\rangle\right).
\end{equation}
Finally, the state Charlie received becomes
\begin{equation}\label{4}
    |\psi_{\rm 1}\rangle=\frac{1}{\sqrt{p_1}}(\alpha a|0\rangle+\beta
    b|1\rangle)
\end{equation}
with the probability $p_1=|a\alpha|^2+|b\beta|^2$ or
\begin{equation}\label{4}
    |\psi_{\rm 2}\rangle=\frac{1}{\sqrt{p_2}}(\beta a|0\rangle+\alpha
    b|1\rangle)
\end{equation}
with the probability $p_2=|a\beta|^2+|b\alpha|^2$. These states are
in accordance with the original state $|\psi\rangle$ only if the
quantum channel is a maximally entangled state, i.e. $\alpha=\beta$.
For the case of  non-maximally entangled channel, there exists the
\textquotedblleft error" in $|\psi_1\rangle$ and $|\psi_2\rangle$.
 These states
can be returned to the original state with certain probability by
performing the generalized measurerment given by Kraus operators:
\addtocounter{equation}{1}
\begin{align}
E_{S1}&=| 0 \rangle\langle 0 |+\frac{\alpha}{\beta}| 1 \rangle\langle 1 |, \tag{\theequation a}\\
E_{F1}&= \sqrt{1-\frac{\alpha^2}{\beta^2}} | 1 \rangle\langle 1
|\tag{\theequation b}
\end{align}
for $|\psi_1\rangle$ and \addtocounter{equation}{1}
\begin{align}
E_{S2}&=\frac{\alpha}{\beta}| 0 \rangle\langle 0 |+| 1 \rangle\langle 1 |, \tag{\theequation a}\\
E_{F2}&= \sqrt{1-\frac{\alpha^2}{\beta^2}} | 0 \rangle\langle 0
|\tag{\theequation b}
\end{align}
for $|\psi_2\rangle$. When $E_S$ is obtained, the qubit ends in its
original state $|\psi\rangle=a|0\rangle+b|1\rangle$. The success
probability in the first teleportation is
\begin{equation}\label{6}
    p=\sum_{i=1}^2p_i\langle\psi_i|E^\dag_{Si}E_{Si}|\psi_i\rangle=2\alpha^2.
\end{equation}
  Next, Charlie teleports the recovered quantum state to Bob by the
similar process. Combining these two teleportations, the total
probability that Bob receives the quantum state $|\psi\rangle$ is
\begin{equation}\label{7}
    P_S=p^2=4\alpha^4.
\end{equation}

However, the above teleportation protocol is not the optimal
strategy. In fact, the third party Charlie does not need to recover
the quantum state to be teleported, but teleports the
\textquotedblleft error state"  to Bob directly. Lastly, Bob
corrects all \textquotedblleft errors" of the quantum state in the
teleportation process. Formally, either Alice and Charlie or Charlie
and Bob do not complete an intact separate teleportation, so we call
it the global multiple teleportation protocol (GMTP).

Let us, thus, assume that Charlie does not correct the
\textquotedblleft error" introduced in the first teleportation, he
only makes a Pauli transformation according to Alice's measurement
outcome, then he
 also performs BM on his two qubits and broadcasts
the measurement outcome to Bob. After making the corresponding Pauli
transformation conditioned on Charlie's measurement outcome, Bob's
qubit will collapse into one of the following states
 \addtocounter{equation}{1}
\begin{align}
& |\phi_{1}\rangle=\frac{1}{\sqrt{p'_1}}(\alpha^2
    a|0\rangle+\beta^2
    b|1\rangle),\tag{\theequation a}\\
& |\phi_{2}\rangle=\frac{1}{\sqrt{p'_2}}(\beta^2
    a|0\rangle+\alpha^2
    b|1\rangle),\tag{\theequation b}\\
& |\phi_{3}\rangle=
    a|0\rangle+b|1\rangle) \tag{\theequation c}
\end{align}
with the probabilities $p'_1=\alpha^4|a|^2+\beta^4|b|^2,$ $
p'_2=\beta^4|a|^2+\alpha^4|b|^2,$ $ p'_3=2\alpha^2\beta^2$
respectively. When the state is in $|\phi_3\rangle$, we do not have
to perform the error correction. It is very joyful to see that the
second teleportation corrects the \textquotedblleft error"
 introduced by the first teleportation. This
effect is called error self-correction. For $|\phi_1\rangle$ and
$|\phi_2\rangle$, one can recover the original state by performing
generalized measurement given by Kraus operators:
\addtocounter{equation}{1}
\begin{align}
E'_{S1}&=| 0 \rangle\langle 0 |+\frac{\alpha^2}{\beta^2}| 1 \rangle\langle 1 |, \tag{\theequation a}\\
E'_{F1}&= \sqrt{1-\frac{\alpha^4}{\beta^4}} | 1 \rangle\langle 1 |
\tag{\theequation b}
\end{align}
and \addtocounter{equation}{1}
\begin{align}
E'_{S2}&=\frac{\alpha^2}{\beta^2}| 0 \rangle\langle 0 |+| 1 \rangle\langle 1 |, \tag{\theequation a}\\
E'_{F2}&= \sqrt{1-\frac{\alpha^4}{\beta^4}} | 0 \rangle\langle 0 |
\tag{\theequation b}
\end{align}
respectively. The total probability of successfully recovering the
original state is
\begin{equation}\label{11}
    P_G(3)=2\alpha^2\beta^2+\sum_{i=1}^2p'_i\langle\phi_i|E'^\dag_{Si}E'_{Si}|\phi_i\rangle=2\alpha^2.
\end{equation}
The ratio of efficiency of GMTP to that of  SMTP
\begin{equation}\label{r}
    P_G(3)/P_S=\frac{1}{2\alpha^2}.
\end{equation}
 We can easily see $P_G/P_S\geq 1$ because of $\alpha\leq\frac{1}{\sqrt{2}}$
. It is obvious that for the maximally entangled channel, the two
protocols are equivalent, but for the partially entangled channel,
GMTP is more efficient than SMTP. Moreover, the less $\alpha$ is,
the more efficient the GMTP is.

It is straightforward to generalize the above two protocols to
arbitrary parties. Let us first discuss the GMTP. Since error
self-correction only appears in the even times Bell-basis
measurements, so here we discuss the $(2N+1)$-party teleportation.
Suppose that Alice 1 wants to teleport a quantum state
$|\psi\rangle=a|0\rangle+b|1\rangle$ to Alice $2N+1$. There is no
direct entanglement resource between them, but they can link through
$2N-1$ intermediaries called Alice 2, Alice 3, $\cdots$, Alice $2N$,
respectively. Two neighboring parties share the partially entangled
state described by Eq.(2). They can complete the task through
cooperative teleportation. After $2N$ Bell-basis measurements and
corresponding Pauli transformations conditioned on previous parties,
the final receiver's qubit will be in one of the states
\begin{equation}\label{12}
       |\phi_i\rangle=\frac{1}{\sqrt{p^G_i}}(\alpha^{2N-i}\beta^i a|0\rangle+\alpha^i\beta^{2N-i}
    b|1\rangle), \\
    \end{equation} with the probability
$C_{2N}^ip_i^G\equiv
C_{2N}^i(\alpha^{2(2N-i)}\beta^{2i}|a|^2+\alpha^{2i}\beta^{2(2N-i)}|b|^2)$,
$i=0, 1, 2, \cdots,  2N$. By correcting the error, the total success
probability is
\begin{equation}\label{p}
    P_{G}(2N+1)=C_{2N}^N\alpha^{2N}\beta^{2N}+2\sum_{i=0}^{N-1}C_{2N}^i\alpha^{2(2N-i)}\beta^{2i}.
\end{equation}
On the other hand, in the case of SMTP, we must perform $2N$
separate teleportations, then the total success probability equals
\begin{equation}\label{s}
    P_S(2N+1)=p^{2N}=2^{2N}\alpha^{4N}.
\end{equation}
 It is easy to verify that $P_G(2N+1)\geq P_S(2N+1)$.

In order to show how the total success probabilities of two
protocols depend on the entanglement of channels for different $N$,
we will choose concurrence $C$ defined by Wootters as a convenient
measure of entanglement \cite{Wootters}. The concurrence varies from
$C = 0$ of a separable state to $C = 1$ of a maximally entangled
state. For a pure  partially entangled state described by Eq. (2),
the concurrence may be expressed explicitly by $C=2|\alpha\beta|$.

In Fig.1, we plot $P_S$ and $P_G$ as the function of concurrence $C$
for different $N$. We can see that both the total success
probabilities of two protocols declines with the decrease of the
entanglement of channels. Moreover, the greater $N$ is, the more
sharper the success probabilities declines. It shows that the
quantum channel with small entanglement will become unpractical with
the increase of $N$.
 Fig.1 also indicates
explicitly that the  GMTP is  more efficient than SMTP. For example,
for the case of $N=10$, the total success probability of GMTP
$P_G\approx 21\%$ while the total success probability of SMTP $P_S$
only attains $0.14\%$ when the concurrence of channels is $C=0.96$.

\begin{figure}[H]
  \begin{center}
    \includegraphics[width=2.7in]{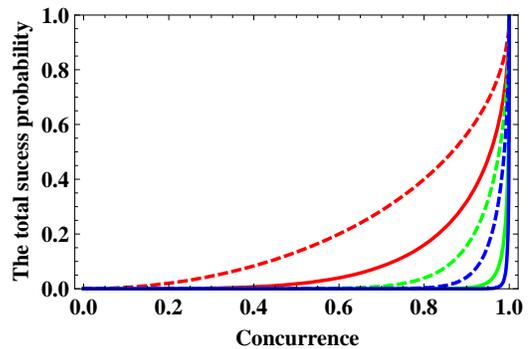}\\
  \caption{The total success probability $P_G$ and $P_S$ versus concurrence $C$ for different $N$
   (Solid line: $P_S$, dashed line: $P_G$).
   From top to bottom, $N$ corresponding takes $1, 5, 10$.}\label{1}
  \end{center}
\end{figure}

The ratio of $P_G$ to $P_S$ as a function of $C$ for different $N$
is illustrated in Fig.2. Here we only take the concurrence from
$0.9$ to $1$ because the small entanglement channels are unpractical
for large $N$.  From Fig.2, we can see that the greater $N$ is, the
larger $P_G/P_S$ is. In other words, the efficiency of GMTP is far
higher than that of SMTP when the steps of teleportation increase.
\begin{figure}[H]
  \begin{center}
    \includegraphics[width=2.7in]{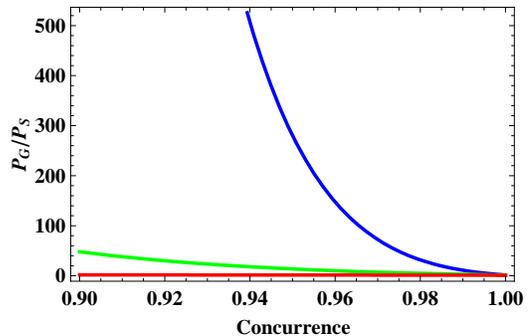}\\
  \caption{The ratio $P_G$ to $P_S$versus concurrence $C$ for different $N$ .
   From bottom to top, $N$ corresponding takes $1, 5, 10$.}\label{1}
  \end{center}
\end{figure}

When the entanglement of the quantum channel is different between
neighboring parties, the circumstance becomes complicated. Here we
only consider the case of three parties .

Alice wants to teleport an unknown quantum state $
    |\psi\rangle=a|0\rangle+b|1\rangle
$ to Bob.  There is no direct entanglement resource between Alice
and Bob, fortunately, Alice and the third party Charlie share a
partially entangled state
\begin{equation}\label{1}
    |\Psi_1\rangle=\alpha_1|00\rangle+\beta_1|11\rangle,
\end{equation}
while  Charlie and Bob share another  entanglement resource
\begin{equation}\label{1}
    |\Psi_2\rangle=\alpha_2|00\rangle+\beta_2|11\rangle,
\end{equation}
where $\alpha_i$ and $\beta_i$ are real numbers and satisfy
$|\alpha_i|\leq |\beta_i|$ and $\alpha_i^2+\beta_i^2=1$. After two
Bell-basis measurements and Pauli operations, the qubit of Bob will
be in one of the following states
\begin{equation}\label{14}
   |\psi_{ij}\rangle=\frac{1}{\sqrt{p_{ij}}}(\alpha_1^{1-i}\alpha_2^{1-j}\beta_1^{i}\beta_2^{j} a|0\rangle +
   \alpha_1^i\alpha_2^j\beta_1^{1-i}\beta_2^{1-j} b|1\rangle)
\end{equation}
with the probabilities
$$
p_{ij}=|\alpha_1^{1-i}\alpha_2^{1-j}\beta_1^{i}\beta_2^{j}
a|^2+|\alpha_1^i\alpha_2^j\beta_1^{1-i}\beta_2^{1-j} b|^2 (i, j=0,
1)
$$
respectively. The qubit can be returned to its original state by
performing the generalized measurement given by Kraus operators:
\addtocounter{equation}{1}
\begin{align}
E_{S}&=| 0 \rangle\langle 0
|+\frac{\alpha_1^{1-i}\alpha_2^{1-j}\beta_1^{i}\beta_2^{j}
}{\alpha_1^i\alpha_2^j\beta_1^{1-i}\beta_2^{1-j}}| 1 \rangle\langle 1 |, \tag{\theequation a}\\
E_{F}&=
\sqrt{1-\frac{|\alpha_1^{1-i}\alpha_2^{1-j}\beta_1^{i}\beta_2^{j}|^2}{|\alpha_1^i\alpha_2^j\beta_1^{1-i}\beta_2^{1-j}|^2}}
| 1 \rangle\langle 1 |\tag{\theequation b}
\end{align}
for $|\alpha_1^{1-i}\alpha_2^{1-j}\beta_1^{i}\beta_2^{j} |\leq
|\alpha_1^i\alpha_2^j\beta_1^{1-i}\beta_2^{1-j} |$. A similar
measurement exists if
$|\alpha_1^{1-i}\alpha_2^{1-j}\beta_1^{i}\beta_2^{j} |\geq
|\alpha_1^i\alpha_2^j\beta_1^{1-i}\beta_2^{1-j} |$.

By tedious but standard calculation we can obtain the success
probability of teleportation
\begin{equation}\label{p}
    P=\min\{2\alpha_1^2, 2\alpha_2^2\}.
\end{equation}
It is an interesting result, the  success probability of
teleportation is completely determined by the channel of less
entanglement. For another channel of more entanglement, its
entanglement does not affect the success probability at all. In
other words, the channel of more entanglement is equivalent to the
maximally entangled channel in the total teleportation process.

 In summary, we have presented two multiple teleportation protocols
via some partially entangled state, the separate multiple
teleportation and the globe multiple teleportation. In the former
protocol, a complete teleportation including error correction is
strictly executed by neighboring parties. However, in the latter
protocol, all errors introduced in the teleportation are corrected
by the final receiver. It has been shown that the property of self
error-correction is a general feature of multiple teleportations,
not being restricted to the scheme proposed in Ref.\cite {s3}. We
also have compared the efficiencies of the two multiple
teleportation protocols and found the globe multiple teleportation
is more efficient than the separate multiple teleportation due to
the property of self error-correction.


\begin{thebibliography}{17}
\bibitem{s1} C.H. Bennett, G. Brassard, C. Cr\'{e}peau, R. Jozsa, A. Peres, and  W.K. Wootters, Phys. Rev. Lett. \textbf{70}, 1895 (1993).
\bibitem{s14} X.B. Wang, T. Hiroshima, A. Tomita, and M. Hayashi,  Phys.
Rep. \textbf{448}, 1 (2007).
\bibitem {s15} G.L. Long, F.G. Deng, C. Wang, X.H. Li, K. Wen, and W.Y.
Wang, Front. Phys. China \textbf{2}, 251 (2007).
\bibitem{s2} D. Gottesman and I.L. Chuang, Nature (London) \textbf{402}, 390 (1999).
\bibitem{s3} E. Knill, R. Laflamme, and G.J. Milburn, Nature (London) \textbf{409}, 46
(2001).
\bibitem{s5} P. Agrawal and A.K. Pati, Phys. Lett. A \textbf{305}, 12 (2002).
\bibitem{s6} G. Gordon and G. Rigolin, Phys. Rev. A \textbf{73}, 042309 (2006).
\bibitem{s7} W.L. Li, C.F. Li, and G.C. Guo, Phys. Rev. A \textbf{61}, 034301 (2000).

\bibitem {s16}F.G. Deng,
C.Y. Li, Y.S. Li, H.Y. Zhou, and Y. Wang, Phys. Rev. A \textbf{72},
022338 (2005).
\bibitem{s9} F.L. Yan
and  D. Wang, Phys. Lett. A \textbf{316}, 297 (2003).
\bibitem{s10} T. Gao, F.L. Yan, and  Y.C. Li, Europhys. Lett.  \textbf{84}, 50001 (2008).
\bibitem{s11} M.Y. Wang and F.L. Yan, Phys. Lett. A \textbf{355}, 94 (2006).
\bibitem{s12} M.Y. Wang, F.L. Yan, T. Gao, and Y.C. Li, International Journal of Quantum Information  \textbf{6}, 201 (2008).
\bibitem{s13} T. Gao, F.L. Yan, and  Z.X. Wang, Quantum Information and Computation  \textbf{4}, 186 (2004).
\bibitem{s4} C.H. Bennett1, G. Brassard, S. Popescu, B. Schumacher, J.A. Smolin,
and W.K. Wootters, Phys. Rev. Lett. \textbf{76}, 722
(1996).
\bibitem{s8} J. Mod{\l}awska and A. Grudka, Phys. Rev. Lett.
\textbf{100}, 110503 (2008).
\bibitem{Wootters} W.K. Wootters, Phys. Rev. Lett.
\textbf{80}, 2245 (1998)

\end{thebibliography}

\end{document}